\documentclass[showpacs,showkeys,twocolumn]{revtex4-1}
\usepackage{bm,graphicx,subfigure}
\begin{document}
\title{Exact Diagonalization Study of Bose-Condensed Gas with Finite-Range\\ Gaussian Interaction}
\author{Mohd. Imran}\email{alimran5ab@gmail.com}
\author{M. A. H. Ahsan}
\affiliation{Department of Physics, Jamia Millia Islamia (Central University), New Delhi 110025, India.}

\begin{abstract} 
We investigate a system of $N$ spinless bosons confined in quasi-two-dimensional harmonic trap with repulsive two-body finite-range Gaussian interaction potential of large $s$-wave scattering length. Exact diagonalization of the Hamiltonian matrix is carried out to obtain the $N$-body ground state as well as low-lying excited states, using Davidson algorithm in beyond lowest-Landau-level approximation. We examine the finite-range effects of the interaction potential on the many-body ground state energy as also the degree of condensation of the Bose-condensed gas. The results obtained indicate that the finite-range Gaussian interaction potential enhances the degree of condensation compared to the zero-range interaction potential. We further analyze the effect of finite-range interaction potential on the breathing mode collective excitation. Our theoretical results may be relevant for experiments currently conducted on quasi-two-dimensional Bose gas with more realistic interaction potential. 

\end{abstract}
\maketitle
\section{Introduction}
\label{intro}
Since the experimental realization of Bose-Einstein condensation (BEC) in harmonically confined ultra-cold alkali atomic vapors \cite{aem95,dma95,bst95}, there has been attempts to examine the role of various energy scales in the physics of many-boson systems \cite{bdz08}.
With recent advances in BEC on optical lattices \cite{mo06} and microchip trap \cite{lsh03}, the few-body systems have attracted particular attention of theorists \cite{blu12}.
The experimental control over parameters such as density, effective dimensionality and the atom-atom interaction strength \cite{ias98}, enables one to examine their effects on the quantum many-body states.
Recent theoretical studies have demonstrated that the breathing mode in a trapped atomic vapor system \cite{pr97,cbr02} is ideally suited as a diagnostic tool for probing quantum many-body effects \cite{moa13}. 
\\
\indent
For dilute gas systems at low-enough temperatures, the interparticle interaction being short-ranged is usually described by zero-range ($\delta$-function) potential \cite{dgp99}. Unfortunately, such an interaction potential in 2D is not self-adjoint \cite{ber98} and cannot be used in a manner analogous to its one dimensional counterpart.
Indeed a regularized 2D contact potential is analytically tractable \cite{far10} but harder to handle numerically.
Further, as the number of particles is increased beyond two, the analytical treatment becomes intractable, leaving one only with a numerical recourse.
\\
\indent 
It has been demonstrated that in order to tackle the limit of zero-range interaction numerically \cite{dka13}, a prohibitively large Hilbert space is required to obtain the ground state and the low-lying excited states. 
For numerical many-body simulations, one usually prefers a smooth, finite-range, model interaction potential. 
This leads us to use the finite-range Gaussian potential as a model interparticle interaction in many-body simulations \cite{cfa09,dka13}, for trapped interacting few-boson system. 
More control over the interparticle interaction with the variation of interaction range and being expandable within finite number of basis functions of the Hilbert space, are few of the advantages of the Gaussian potential over the usual $\delta$-function potential.
\\
\indent 
In this note, we present an exact diagonalization study of quasi-two-dimensional system of $N$ spinless bosons interacting via finite-range Gaussian potential. 
The exact diagonalization of the many-body Hamiltonian matrix is performed using beyond lowest-Landau-level approximation, to obtain the low-lying energy spectrum of ultra-cold Bose gas in a harmonic trap.
Our approach involves the inclusion of lower as well as higher Landau levels with single-particle angular momentum $m$ of either sign for construction of $N$-body basis function \cite{ahs01}.
The results obtained demonstrate that the use of (repulsive) particle-particle Gaussian interaction has a dramatic effect on various properties of the Bose-condensed gas. 
\\
\indent
This paper is organized as follows.
In Sec.~\ref{model}, we describe the model Hamiltonian for Bose gas with repulsive finite-range Gaussian interaction potential, confined in a quasi-two-dimension harmonic trap.
We then introduce the single-particle reduced density matrix to delineate the criterion for the existence of Bose-Einstein condensate. We offer a justification for the use of Gaussian interaction potential in our exact diagonalization calculation, instead of the usual contact ($\delta$-function) potential. 
In Sec.~\ref{results}, we present the exact results for a system of $N$ bosons, to explore the finite-range effect of two-body Gaussian interaction potential, on the many-body ground state as well as on the first breathing mode collective excitation of the system.
Finally, in Sec.~\ref{conc}, we summarize our results and conclusions of the present study.
\section{The Model}
\label{model}
We consider a system of $N$ interacting spinless bosons each of mass $M$, trapped in a harmonic potential $V({\bf r})= {\frac{1}{2}}M\left(\omega_{\perp }^{2} {r}^{2}_{\perp}+\omega_{z}^{2} {z}^{2}\right)$. 
Here, $r_{\perp}=\sqrt{x^{2}+y^{2}}$ is the radial distance of a particle from the trap center; $\omega_{\perp}$ and $\omega_{z}$ are the radial and axial frequencies respectively, of the harmonic confinement.
The trapping potential $V\left({\bf r}\right)$ is assumed to be highly anisotropic with $\lambda_{z}\equiv \omega_{z} / \omega_{\perp}\gg 1$ so that the many-body dynamics along $z$-axis is frozen. The system is thus effectively quasi-two-dimensional (quasi-2D) with $x$-$y$ rotational symmetry.
Choosing $\hbar \omega_{\perp}$ as the unit of energy and $a_{\perp} = \sqrt{\hbar/{M \omega_{\perp}}}$ the corresponding unit length, the many-body Hamiltonian in dimensionless form is given by
\begin{equation}
H = \sum_{j=1}^{N} \left[-\frac{1}{2} \bm{\nabla}^{2}_{j} + \frac{1}{2} {\bf r}_{j}^{2} \right] 
+ \frac{1}{2} \sum_{i\neq j}^{N} U \left({\bf r}_{i},{\bf r}_{j}\right)
\label{mbh}
\end{equation} 
The first two terms in the Hamiltonian~(\ref{mbh}) correspond to the kinetic and potential energies respectively, and the third term arises from the particle-particle interaction.
At low-enough temperatures, the interparticle interaction $U\left({\bf r}_{i},{\bf r}_{j}\right)$ is described by the Gaussian potential 
\begin{equation}
U \left({\bf r}_{i},{\bf r}_{j}\right) = \frac{\mbox{g}_{2}} {2\pi{\sigma^{2}}}
\exp{\left[ -\frac{\left(r_{\perp i}-r_{\perp j}\right)^{2}}{2\sigma^{2}} \right]} 
\delta \left(z_{i}-z_{j}\right)
\label{gip}
\end{equation}
with $\sigma$ (scaled by $a_{\perp}$) being the effective range 
of Gaussian interaction.
The dimensionless parameter $\mbox{g}_{2}=4\pi {a_{s}}/{a_{\perp}}$ measures the strength of the two-body interaction with $a_{s}$ being the $s$-wave scattering length for particle-particle collision. 
We assume that the scattering length is positive $\left(a_{s}>0\right)$ so that the effective finite-range interaction is repulsive. 
The above Gaussian interaction potential~(\ref{gip}) is expandable within a 
finite number of single-particle basis functions and hence computationally more feasible \cite{cfa09,dka13}, compared to the zero-range $\delta$-function potential.  
In the limit $ \sigma \rightarrow 0$, the normalized Gaussian potential in Eq.~(\ref{gip}) reduces to the zero-range contact potential $\mbox{g}_{2}\delta\left({\bf r}_{i}-{\bf r}_{j}\right)$ \cite{dgp99}.
\\
\indent
It is to be noted that for a many-body system under consideration here, the  characteristic energy scale for the interaction is determined by the  dimensionless parameter $\left(Na_{s}/a_{\perp}\right)$. 
Owing to the increasing dimensionality of the Hilbert space with $N$, making computation impractical, we vary $a_{s}$ so as to achieve a suitable value of  $\left(N a_{s}/a_{\perp}\right)$ relevant to experimental situation \cite{dgp99}.
To obtain the eigenenergies $E_{k}$ and the corresponding eigenstates $\Psi_{k}$ of the $N$-boson system, we employ exact diagonalization of the Hamiltonian matrix using Davidson algorithm \cite{dav75} with inclusion of lower as well as higher Landau levels in the construction of $N$-body basis states \cite{ahs01}.
The index $k$ labels the variationally obtained $k$th state and the corresponding many-body wavefunction $\Psi_{k}$ of the system.
\\
\indent
The $N$-body ground state wavefunction $\Psi_{0}({\bf r}_{1},{\bf r}_{2},\dots,{\bf r}_{N})$ is assumed to be normalized; one can then determine the single-particle reduced density matrix $\rho_{1}({\bf r},{\bf r}^{\prime})$, by integrating out the degrees of freedom of $N-1$ particles. 
Thus
\begin{eqnarray}
\rho_{1}({\bf r},{\bf r}^{\prime}) &=&\int \int \dots \int d{\bf r}_{2}\ d{\bf r}_{3}\dots d{\bf r}_{{N}}\nonumber \\
&&\times \ \Psi_{0}^{\ast} ({\bf r},{\bf r}_{2},{\bf r}_{3} \dots,{\bf r}_{{N}})\ \Psi_{0} ({\bf r}^{\prime},{\bf r}_{2},{\bf r}_{3},\dots,{\bf r}_{{N}}) \nonumber \\
&\equiv & \sum_{{\bf n},{\bf n}^{\prime}}\rho_{_{{\bf n},{\bf n}^{\prime}}}u^{\ast}_{\bf n}\left({\bf r}\right)u_{{\bf n}^{\prime }}\left({\bf r}^{\prime }\right).  
\end{eqnarray}
The above expression is written in terms of single-particle basis functions $u_{\bf n}\left({\bf r}\right)$ with quantum number ${\bf n}\equiv \left(n,m,n_{z}\right)$.
Being hermitian, this can be diagonalized to give 
\begin{equation}
\rho_{1} \left({\bf r},{\bf r}^{\prime}\right) = \sum_{\mu }\lambda_{\mu } \ \chi^{\ast}_{\mu }\left({\bf r}\right)
\chi_{\mu } \left({\bf r}^{\prime }\right),
\label{spd}
\end{equation} 
where $
\chi_{\mu }\left({\bf r}\right) \equiv \sum_{\bf n} c^{\mu }_{\bf n} \, {u}_{\bf n}\left({\bf r}\right)$ and $\sum_{\mu }\lambda_{\mu }=1$
with $ 1\geq\lambda_{1}\geq\lambda_{2}\geq\cdots\lambda_{\mu }\geq\cdots \geq 0 $.
The $\left\{\lambda_{\mu }\right\}$ are the eigenvalues, ordered as above, and $\left\{ \chi_{\mu }\left({\bf r}\right)\right\}$ the corresponding eigenvectors of the single-particle reduced density matrix $\rho_{1}\left({\bf r},{\bf r}^{\prime }\right)$; each $\mu$ defines a fraction of the Bose-Einstein condensate.
\section{Results and Discussion}
\label{results}
We consider a system of $N=10$ bosonic atoms of $^{87}$Rb in a quasi-2D harmonic trap with confining frequency ${\omega}_{\perp}=2\pi \times 220$ Hz and the $z$-asymmetry parameter $\lambda_{z}\equiv {\omega_{z}}/{\omega_{\perp}}=\sqrt{8}$. 
The condensate has extension $a_{z}=\sqrt{\hbar/M\omega_{z}}$ in the $z$-direction and its dynamics along this axis is taken to be completely frozen. 
Recent advancements in atomic physics have made it possible to tune the low-energy atom-atom scattering length in ultra-cold atomic vapors using Feshbach resonance \cite{ias98}. 
Accordingly in the calculations presented here, the parameters of the two-body interaction potential~(\ref{gip}) have been chosen as the range $0 \leq \sigma \leq 1$ (in units of $a_{\perp}$) and the $s$-wave scattering length $a_{s}=1000a_{0}$ where, $~a_{0}=0.05292~nm$ is the Bohr radius. 
The corresponding value of the dimensionless interaction parameter $\mbox{g}_{2}={4\pi a_{s}}/{a_{\perp}}$ turns out to be $0.9151$.  
From now on we fix the value of dimensionless interaction parameter $\mbox{g}_{2}=0.9151$ and vary the value of Gaussian width $\sigma$ from $0$ to $1$ (in units of $a_{\perp}$).
Our results obtained through exact diagonalization with repulsive Gaussian interaction potential allow us to study the effect of the range $\sigma$ of interparticle interaction on various properties of the Bose-condensed gas, as discussed in the following.
\subsection{Ground State: Energy and Degree of Condensation}
In this subsection, we examine the behavior of ground state energy and degree of condensation with the range of interaction varied over $0\le \sigma \le 1$ for the repulsive finite-range Gaussian interaction potential~(\ref{gip}). 
\begin{figure}[!htb]
\centering
\includegraphics[width=0.9\linewidth]{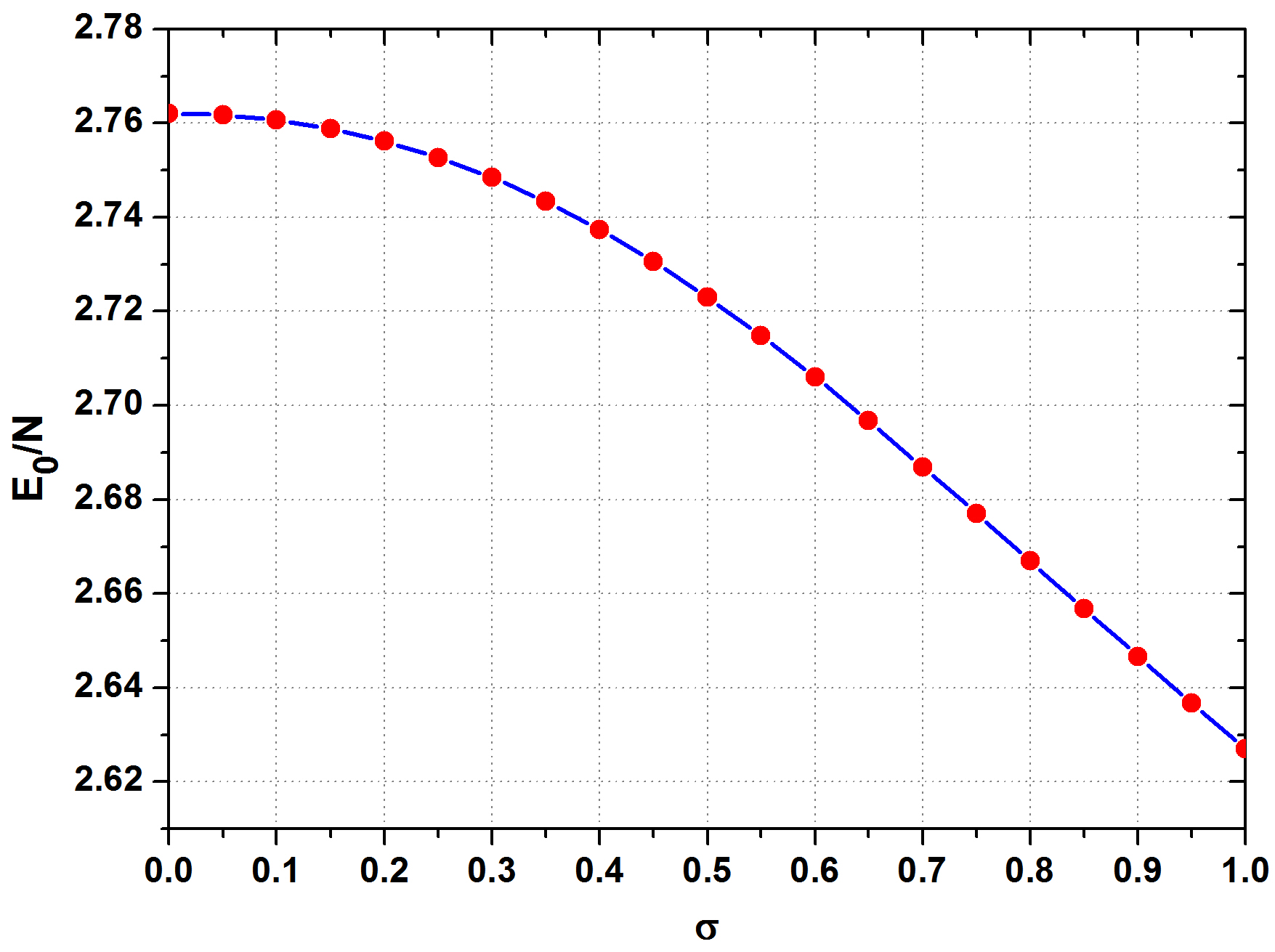}
\caption{\label{gse} The ground state energy per particle $\left(E_{0}/N\right)$ {\it versus} the range $\sigma$ (in units of $a_{\perp}$) of Gaussian interaction potential with interaction parameter $\mbox{g}_{2}=0.9151$ for a system of $N=10$ trapped bosons. The limiting case $\sigma \rightarrow 0$ corresponds to contact ($\delta$-function) interaction potential.}
\end{figure}
\\
\indent
In Fig.~\ref{gse}, we plot the ground state energy per particle $\left(E_{0}/N\right)$ as a function of the range $\sigma$ of the repulsive interaction potential with $\mbox{g}_{2}=0.9151$ for $N=10$ bosons.
As is observed from the figure that the ground state energy changes very little for small values of $\sigma$, but exhibits a monotonic decrease (enhancing the phase rigidity of the condensate) as $\sigma$ is further increased over the range $0 < \sigma \le 1$, in units of $a_{\perp}$. For larger values of $\sigma \gg 1$ (not shown in the figure), the ground state energy is found to approach the non-interacting value.
\\
\indent
To further examine the effect of $\sigma$ on quantum mechanical phase coherence of the Bose-condensed gas, we calculate the single-particle density matrix defined in Eq.~(\ref{spd}), where the eigenvalues have been normalized to 1.
It is to be noted that the usual definition of condensation for a macroscopic system, given by the largest eigenvalue $\lambda_{1}$ of the single-particle reduced density matrix, is not appropriate for systems with small number of particles being studied here.
For example, in the absence of condensation, there is no macroscopic occupation of a single quantum state and all levels are equally occupied, such a definition would imply a condensate fraction with small magnitude.
To circumvent this situation, one introduces a quantity which is sensitive to the loss of macroscopic occupation called the degree of condensation defined as
\begin{equation}
C_{d} =\lambda_{1}-\bar{\lambda}
\label{doc}
\end{equation}
where $\bar{\lambda}=\frac{1}{p-1}\sum_{\mu=2}^{p} \lambda_{\mu}$ is the mean of the rest of eigenvalues. 
It can be seen that the degree of condensation, defined in Eq.~(\ref{doc}), approaches zero for equal eigenvalues, as one would expect.
\begin{figure}[!htb]
\centering
\includegraphics[width=0.9\linewidth]{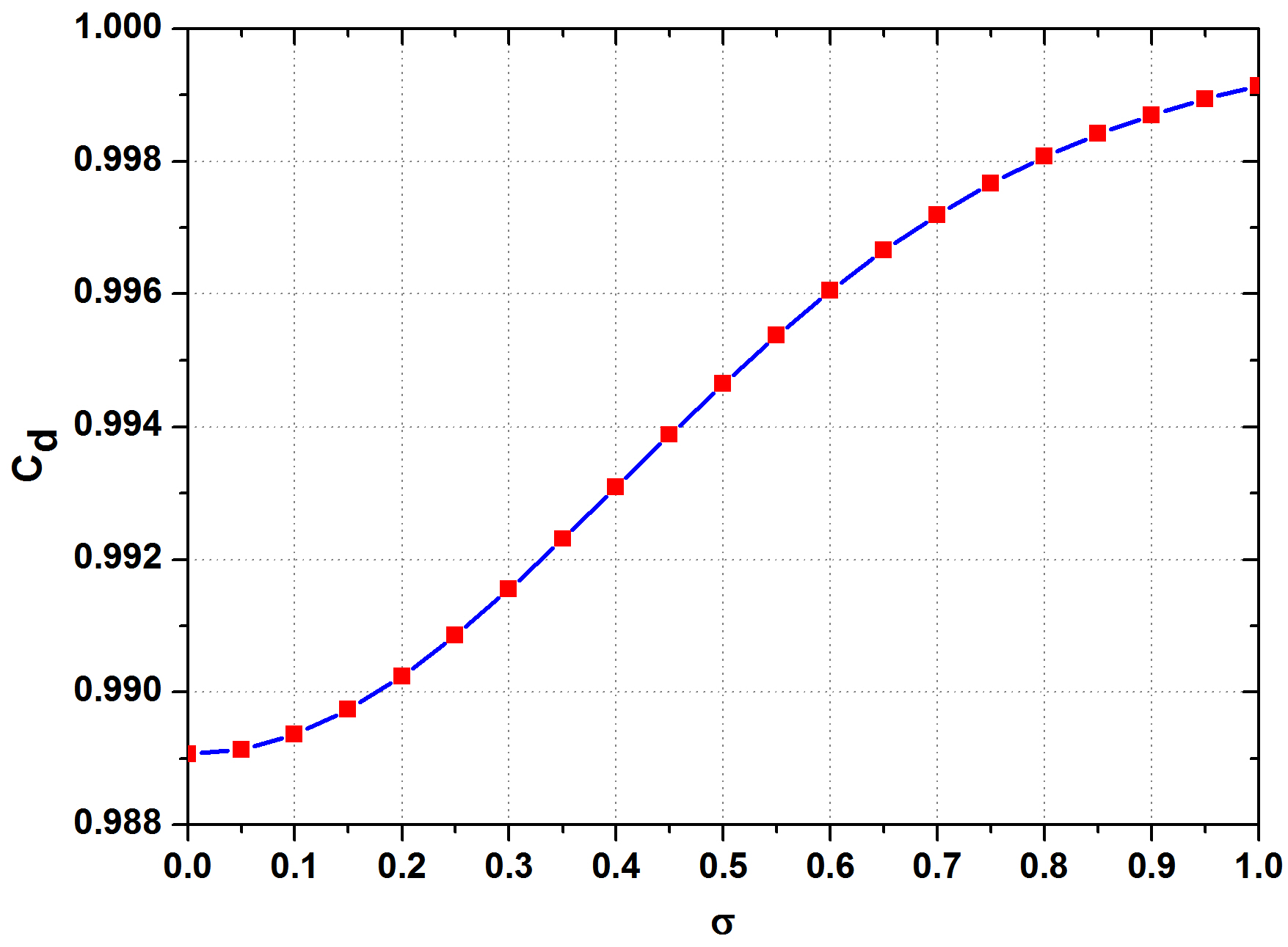}
\caption{\label{gsc} The degree of condensation $C_{d}$ of the ground state as a function of interaction range $\sigma$ (in units of $a_{\perp}$) with interaction parameter $\mbox{g}_{2}=0.9151$ for a system of $N=10$ trapped bosons.}
\end{figure}
\\
\indent
In Fig.~\ref{gsc}, we present the variation of degree of condensation $C_{d}$ with interaction range $\sigma$ for a system of $N=10$ bosons. 
It is seen from the figure that the degree of condensation increases with increase in $\sigma$ and the variation becomes nearly linear for larger values of $\sigma \le 1$. 
The explanation for such a behavior is that as the range $\sigma$ of the Gaussian interaction potential increases, the interparticle potential begin to overlap leading to increase in many-body effects. This increased many-body effect compared to the zero-range ($\delta$-function) potential leads to an enhanced phase coherence, reflected in increased degree of condensation. 
\subsection{Excited State: First Breathing Mode}
It was pointed out by Pitaevskii and Rosch \cite{pr97} that a purely 2D harmonically confined Bose gas interacting via a zero-range ($\delta$-function) interaction potential, exhibits breathing modes arising from $SO(2,1)$ symmetry. 
This underlying symmetry leads to an energy spectrum, generating eigenmodes with energy spacing of $2\hbar \omega_{\perp}$ between two adjacent breathing modes.
Although in the present work, the finite-range interaction potential has been used in place of the zero-range interaction, the feature of $2\hbar \omega_{\perp}$ spacing in the energy eigenspectrum, is seen to persist for certain range of values of $\sigma $ for the Gaussian interaction potential~(\ref{gip}).
For instance, the energy spacing between the excited state $E_{1}$ and the ground state $E_{0}$ for $N=10$ bosons is $E_{BM}=E_{1}-E_{0}=2.0111$ (in units of $\hbar \omega_{\perp}$) with $\sigma=0.1$ and $\mbox{g}_{2}=0.9151$ (corresponding to $a_{s}=1000~a_{0}$). 
\begin{figure}[!htb]
\centering
\includegraphics[width=0.9\linewidth]{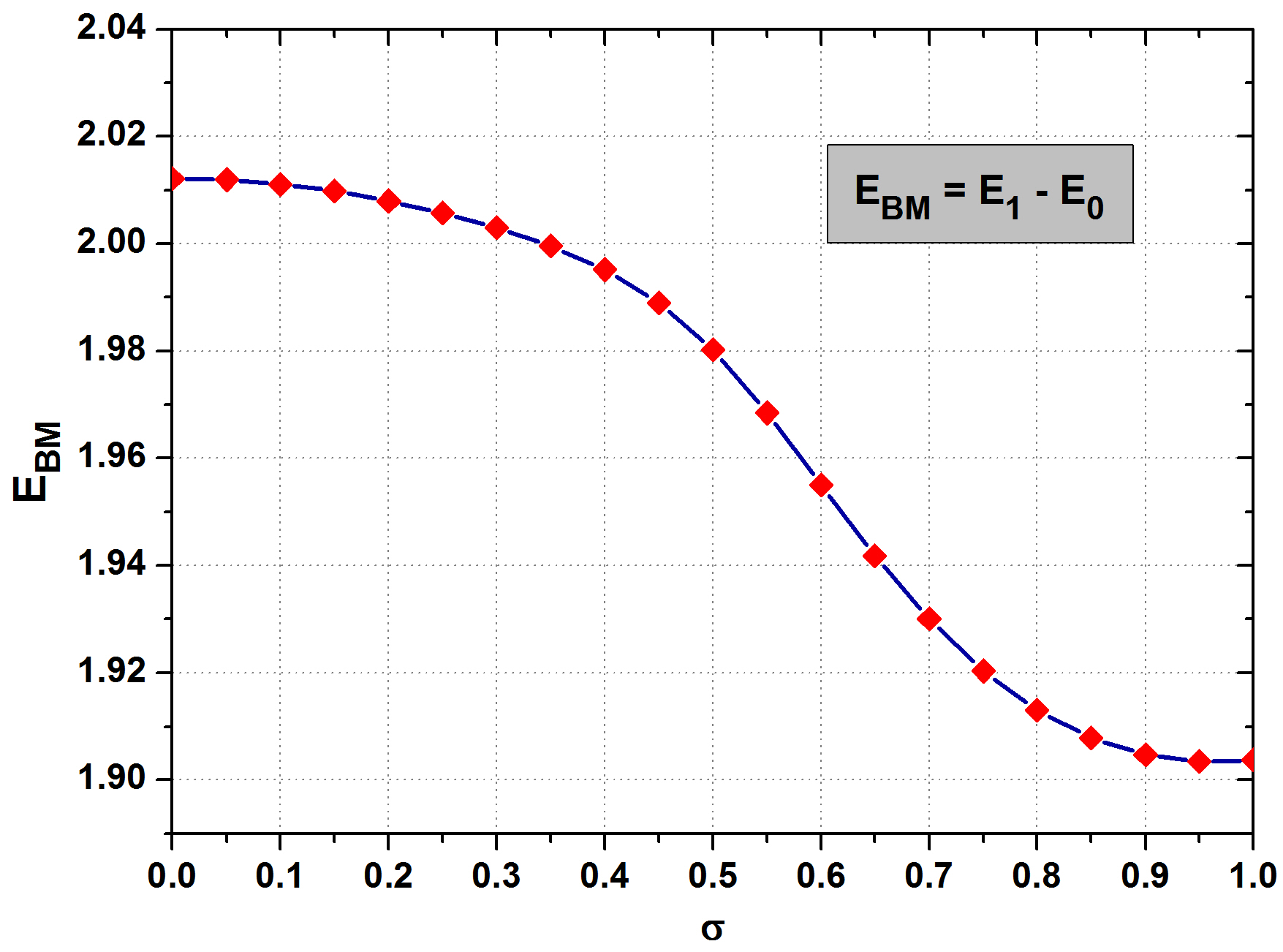}
\caption{\label{bme} For $N=10$ bosons, the energy $E_{BM}$ (in units of $\hbar \omega_{\perp}$) of the first breathing mode {\it versus} the interaction range $\sigma$ (in units of $a_{\perp}$) of the Gaussian potential with fixed value of the interaction parameter $\mbox{g}_{2}=0.9151$ (parametrized by $a_{s}$). The energy value of breathing mode is $E_{BM}=E_{1}-E_{0}$, where $E_{1}$ is the energy of excited state and $E_{0}$ is the energy of ground state of the system.}
\end{figure}
\\
\indent 
Next, we examine the effect of $\sigma $ on the eigenenergy of the first breathing mode and observe that the value of the eigenmode $E_{BM}=E_{1}-E_{0}$ deviates from $2\hbar \omega_{\perp}$ which is the value strictly for $\delta$-function interaction ($\sigma =0$) in 2D. 
Figure~\ref{bme} presents the variation of $E_{BM}$ (in units of $\hbar \omega_{\perp}$) with the interaction range $\sigma$ of the Gaussian potential with $\mbox{g}_{2}=0.9151$ for a system of $N=10$ bosons. 
It is observed from the figure that with increasing $\sigma$, the repulsive Gaussian interaction, in general, lowers the value of the first breathing mode $E_{BM}$ from $2\hbar\omega_{\perp}$. For small values of interaction range $\sigma < 0.4$, the value of $E_{BM}$ stays close to $2\hbar \omega_{\perp}$. 
In the interval $0.4 < \sigma < 0.85$, the finite-range effect of the interaction potential is prominently seen and the values of $E_{BM}$ deviate appreciably from $2\hbar \omega_{\perp}$.
However, it is observed that when $\sigma$ approaches the system size $a_{\perp}$, the value of $E_{BM}$ saturates to a value appreciably less than $2\hbar \omega_{\perp}$.
In other words, with increasing $\sigma$ as the Gaussian interaction potential deviates from the $\delta$-function potential, the breathing mode value exhibits a corresponding deviation from the value $2 \hbar \omega_{\perp}$ for a strictly 2D zero-range interaction potential.
\section{Conclusion}
\label{conc} 
In the present study, we have studied a system of $N=10$, harmonically confined bosons in quasi-2D, interacting via  finite-range Gaussian potential, through exact diagonalization in beyond lowest-Landau-level approximation to analyze its ground state and the first breathing mode.
The low-lying energy eigenstates were obtained as the width $\sigma$ of the repulsive Gaussian interaction potential is varied over the range from zero (corresponding to $\delta$-function potential) to system size.
We studied the dependence of ground state energy, the first breathing mode and the degree of condensation on the range $\sigma$ of the interaction potential.
It was observed that the ground state energy decreases as the interaction range is increased.
It was also observed that increase in the width of Gaussian interaction potential monotonously enhances the degree of condensation compared to the zero-range interaction potential.
Further, the range $\sigma$ of the interaction potential also affects the energy of the first breathing mode (collective excitation).
With increasing width $\sigma$, as the Gaussian interaction potential deviates from the $\delta$-function potential, the breathing mode value exhibits a corresponding deviation from the value $2 \hbar \omega_{\perp}$ for a strictly 2D zero-range interaction potential.
The results of the present work suggest the suitability of finite-range Gaussian interaction potential as model of the two-body interaction and may therefore profitably be employed in simulation studies of quasi-2D confined interacting bosons.

\end{document}